\documentclass[%
 reprint,
amsmath,
 amssymb,
 aps,
 prl,
floatfix,
]{revtex4-2}
\usepackage{graphicx}
\usepackage{dcolumn}
\usepackage{bm}

\usepackage{hyperref}

\begin{document}

\preprint{APS/123-QED}

\title{Turbulence in Magnetic Reconnection Jets from Injection to Sub-Ion Scales}

\author{Louis Richard}
 \email{louis.richard@irfu.se}
\affiliation{%
Swedish Institute of Space Physics, Uppsala, Sweden}
\affiliation{Department of Physics and Astronomy, Space and Plasma Physics, Uppsala University, Sweden 
}%

\author{Luca Sorriso-Valvo}
\affiliation{
CNR/ISTP – Istituto per la Scienza e la Tecnologia dei Plasmi, Bari, Italy
}
\affiliation{
Space and Plasma Physics, School of Electrical Engineering and Computer Science, KTH Royal Institute of Technology, Stockholm, Sweden
}%
\affiliation{
Swedish Institute of Space Physics, Uppsala, Sweden
}%

\author{Emiliya Yordanova}
\affiliation{
Swedish Institute of Space Physics, Uppsala, Sweden
}%

\author{Daniel B. Graham}
\affiliation{
Swedish Institute of Space Physics, Uppsala, Sweden
}%

\author{Yuri V. Khotyaintsev}
\affiliation{
Swedish Institute of Space Physics, Uppsala, Sweden
}%

\date{\today}

\begin{abstract}
We investigate turbulence in magnetic reconnection jets in the Earth’s magnetotail using data from the Magnetospheric Multiscale spacecraft. We show that signatures of a limited inertial range are observed in many reconnection jets. The observed turbulence develops on the time scale of a few ion gyroperiods, resulting in intermittent multifractal energy cascade from the characteristic scale of the jet down to the ion scales. We show that at sub-ion scales, the fluctuations are close to mono-fractal and predominantly kinetic Alfv\'en waves. The observed energy transfer rate across the inertial range is $\sim 10^8~\mathrm{J}~\mathrm{kg}^{-1}~\mathrm{s}^{-1}$, which is the largest reported for space plasmas so far. 
\end{abstract}

\maketitle

The interplay between the two ubiquitous phenomena, magnetic reconnection and turbulence, is a long-standing problem in collisionless plasmas~\cite{lazarian_turbulent_2015}. Magnetic reconnection is a process that provides energization and acceleration of plasma through explosive topological reconfiguration of the magnetic field~\cite{biskamp_magnetic_2000,yamada_magnetic_2010}. It is responsible for the generation of fast plasma flows (jets) as observed, for example, in solar flares~\cite{masuda_loop-top_1994}, black hole flares~\cite{ripperda_black_2022} and planetary magnetotails~\cite{yamada_magnetic_2010}. In the reconnection region, turbulence and wave growth due to kinetic processes can, in turn, affect the dynamics of the magnetic reconnection~\cite{khotyaintsev_collisionless_2019}. On the other hand, turbulence is a universal process that transfers kinetic and magnetic energy from large injection scales to small scales through an energy cascade produced by non-linear interactions among fluctuations~\cite{biskamp_magnetohydrodynamic_2003,frisch_turbulence_1995}. If the turbulence is fully developed, such energy transfer is globally scale invariant over a range of scales, called the inertial range, where large-scale forcing and small-scale dissipation can be neglected. This produces power-law scaling of statistical quantities, such as the power spectral density and the moments of the scale-dependent fluctuations~\cite{frisch_turbulence_1995}. In addition, spatial inhomogeneity of the energy transfer results in intermittency, i.e., formation of spatially concentrated structures such as current sheets and vortices~\cite{karimabadi_coherent_2013} where dissipation occurs~\cite{matthaeus_turbulence_2021}.

Numerical simulations show that turbulence develops in reconnection jets, resulting in the formation of secondary reconnection sites~\cite{lapenta_secondary_2015} and intermittent magnetic field fluctuations at kinetic scales (smaller than the ion gyroscale)~\cite{leonardis_identification_2013,pucci_properties_2017}. \textit{In-situ} spacecraft observations in reconnection jets suggest development of turbulence~\cite{voros_bursty_2006,huang_observations_2012,osman_multi-spacecraft_2015,jin_characteristics_2022}, forming current sheets where energy is dissipated~\cite{osman_multi-spacecraft_2015,fu_intermittent_2017}. The interaction of the particles with the turbulence-generated secondary magnetic flux ropes and other spatially concentrated structures provides efficient particle heating and acceleration through the Fermi mechanism~\cite{dahlin_role_2017,li_formation_2019,zhang_efficient_2021,arnold_electron_2021,dalena_test-particle_2014,zhdankin_electron_2019,lemoine_first-principles_2022}. Therefore, a complete description of the turbulent energy transfer from injection to sub-ion scales is crucial to understand the energization of the content of collisionless plasma jets. However, the transient nature of reconnection jets yields short samples of \textit{in-situ} measurements, and thus, it is difficult to obtain a meaningful statistical description of the fluctuations~\cite{dudok_de_wit_can_2004}. As a result, the complex interplay between magnetic reconnection and turbulence in reconnection jets remains unclear. 

In this Letter, we use data from the Magnetospheric Multiscale (MMS) spacecraft \cite{burch_magnetospheric_2016} in the terrestrial magnetosphere to investigate turbulence in reconnection jets. We study 330 plasma jets in the plasma sheet of the Earth's magnetotail ($\beta_i \geq 0.5$, where $\beta_i = 2\mu_0 n_i k_B T_i / B^2$, $n_i$ is the ion number density, $T_i$ the ion temperature, and $B$ the magnetic field magnitude)~\cite[see Ref.][for a detailed description of the data]{richard_are_2022}.  
The jets are observed in the mid-tail, $-15~R_E \geq X_{GSM} \geq -25~R_E$ in geocentric solar magnetospheric (GSM) coordinates; $R_E\approx 6371~\mathrm{km}$ is the Earth’s radius. Magnetic reconnection is the primary driver of fast plasma flows in this part of the magnetotail~\cite{angelopoulos_tail_2008,angelopoulos_electromagnetic_2013}. The statistical location of the reconnection X-line for our dataset is $X_{GSM}\approx -25~R_E$~\cite{richard_are_2022}. Other mechanisms to generate jets (e.g., kinetic ballooning/interchange~\cite{pritchett_kinetic_2010,panov_kinetic_2012}) would be effective only in the near-Earth tail, $X_{GSM}\geq -15~R_E$, which is outside of the region we study. Therefore, we assume that the observed jets are generated by reconnection. We use magnetic field measurements from MMS's FGM instrument \cite{russell_magnetospheric_2016} and electric field measurements from the EDP instrument \cite{lindqvist_spin-plane_2016,ergun_axial_2016}. The moments of the ion and electron velocity distributions are measured by the FPI instrument \cite{pollock_fast_2016} with corrections removing low-energy photo-electrons and a background ion population to account for penetrating radiation \cite{gershman_systematic_2019}.

\begin{figure}
\includegraphics[width=\linewidth]{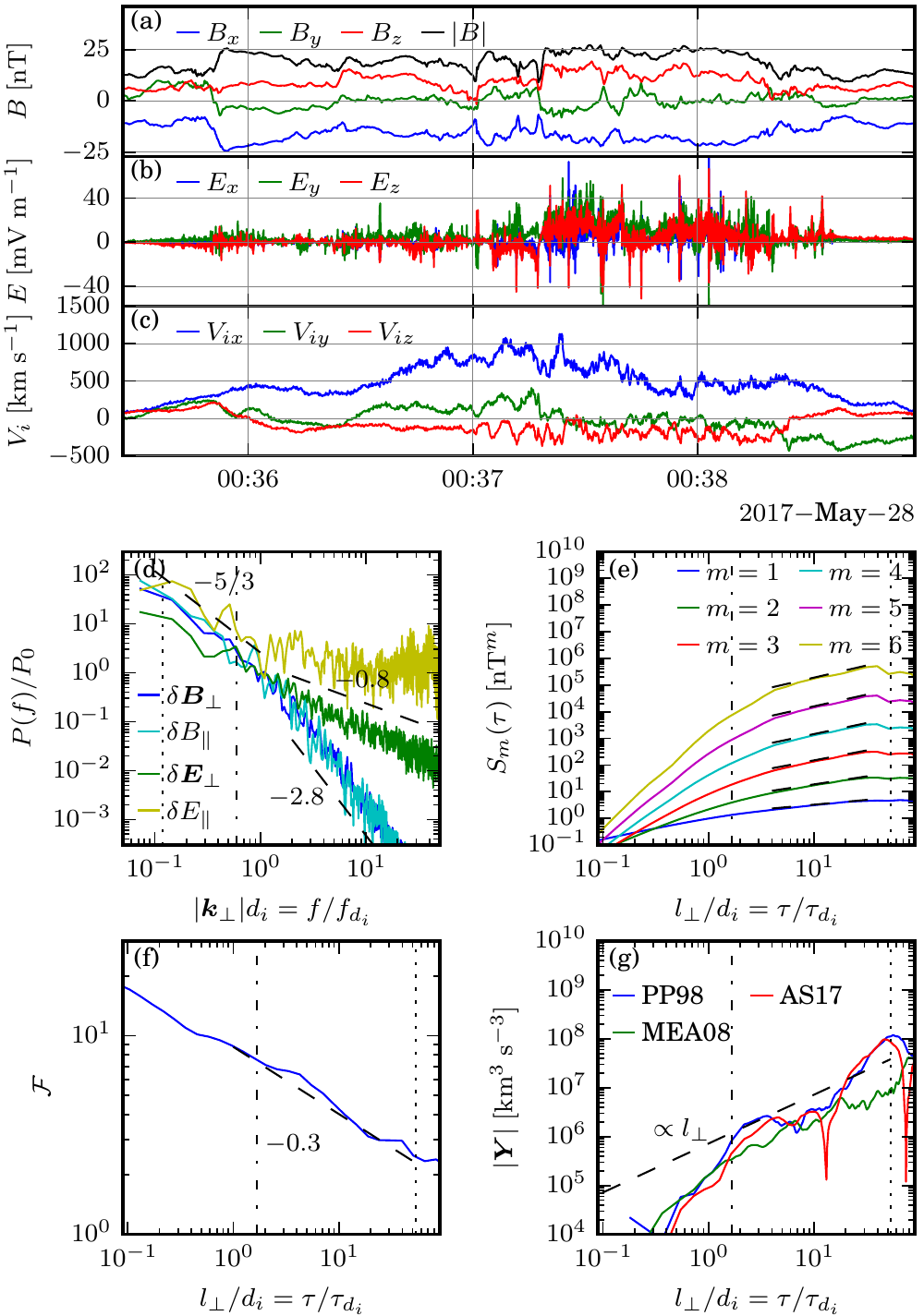}
\caption{\label{fig:example}Example of a reconnection jet with signatures of fully developed turbulence. (a) Magnetic field, (b) electric field, 
and (c) ion bulk velocity in GSM coordinates; (d) power spectral density of the electromagnetic fields normalized to $P_0=P(f_{d_i})$, where $f_{d_i}= V /2\pi d_i$; (e) structure functions and (f) flatness of the magnetic field; (g) energy flux $|\bm{Y}|$. The dotted lines in panels (d)-(g) indicate the correlation scale $l_c$. The dashed-dotted lines in panels (d) and (e)-(g) indicate $|\bm{k}_\perp|\rho_i=1$ and $l_\perp=\rho_i$, respectively. The dashed lines in panels (d)-(g) are reference power laws.}
\end{figure}

Figure \ref{fig:example} presents an example of a fast ($|\bm{V}_i|\geq300~\mathrm{km}~\mathrm{s}^{-1}$ with $\bm{V}_i$ the ion bulk velocity) Earthward jet [Fig.~\ref{fig:example}c]. We see enhanced fluctuations in the magnetic field $\bm{B}$ and electric field $\bm{E}$ [Fig.~\ref{fig:example}a-b]. The correlation scale, i.e., the size of the energy-containing eddies~\cite{matthaeus_evolution_1994,batchelor_theory_1953}, is $l_c=53 d_i=3.7~R_E$, where $d_i=\sqrt{m_i/n_i e^2 \mu_0}$ is the ion inertial length (see Supplemental Material~\cite{supplemental_material}). Here, we use temporal-to-spatial scale equivalence $l_\perp= V \tau$, with $V=\left \langle|\bm{V}_i|\right \rangle$, after verifying the validity of the Taylor hypothesis of frozen-in-flow fluctuations~\cite{taylor_spectrum_1938,howes_validity_2014} and the assumption of anisotropic fluctuations (see Supplemental Material~\cite{supplemental_material}). 
The power spectra of the electromagnetic fluctuations [Fig.~\ref{fig:example}d] exhibit a Kolmogorov-like power-law scaling, $|\bm{k}_\perp|^{-5/3}$~\cite{kolmogorov_local_1941} in a range spanning from the energy injection scale, estimated as the correlation scale $l_c$, down to the ion gyroscale $\rho_i=\sqrt{\beta_i}d_i\approx 619~\mathrm{km}$.  At sub-ion scales, the magnetic field spectrum steepens to $|\bm{k}_\perp|^{-2.8}$ while the electric field spectrum rises to $|\bm{k}_\perp|^{-0.8}$, due to the contribution of the Hall term in the generalized Ohm's law at the ion kinetic scales~\cite{matteini_electric_2017,stawarz_comparative_2021}. The observed Kolmogorov spectrum at large scales ($l_c \geq l_\perp \geq \rho_i$) suggests global scale-invariant energy transfer across these scales.

In fully developed turbulence, the power spectrum (equivalent to the second-order moment of the fluctuations) is not sufficient to describe the fluctuations due to the intermittency~\cite{paladin_anomalous_1987}. Hence, we compute the structure functions of the magnetic field $S_{m}(\tau)=\left \langle |\Delta \bm{B}(\tau)|^m \right \rangle = \left \langle | \bm{B}(t+\tau) - \bm{B}(t)|^m \right \rangle$ with $\tau$ the time scale and $\left \langle \cdot \right \rangle$ the ensemble time average, having verified ergodicity and statistical convergence (see Supplemental Material~\cite{supplemental_material}). We observe a power-law scaling $S_m(\tau)\propto \tau^{\zeta(m)}$ at large scales [Fig.~\ref{fig:example}e], which confirms the global scale-invariant nature of the fluctuations. In addition, the flatness $\mathcal{F}(\tau)=S_4(\tau)/S_2^2(\tau)$, which measures the wings of the distribution of $\Delta \bm{B}(\tau)$, i.e., the occurrence of large gradients, is monotonically increasing as the scale decreases [Fig.~\ref{fig:example}f] indicating intermittency \cite{frisch_turbulence_1995}. This provides evidence for spatially inhomogeneous energy transfer at large scales. 

Knowledge of the energy transfer rate by the turbulence cascade across the scales is crucial to understanding the energy budget in the reconnection jets. We estimate the energy cascade rate using the third-order law for  three-dimensional single-fluid magnetohydrodynamic (MHD) turbulence under the assumption of scale separation between injection and dissipation, homogeneity, and time stationarity of the fluctuations~\cite{politano_von_1998,macbride_turbulent_2008,andres_alternative_2017},

\begin{equation}
    \label{eq:exact-law}
     -2\varepsilon = \frac{1}{2}\nabla_l \cdot \bm{Y} + S,
\end{equation}
\noindent
where $\varepsilon$ is the energy cascade rate, $\bm{Y}$ is the energy flux, and $S$ is the energy source term, which is commonly assumed to be negligible compared with the flux terms~\cite{andres_energy_2019}. 
Using various assumptions, Eq.~\ref{eq:exact-law} can be simplified to formulations which can be applied to spacecraft measurements. 
We employ the anisotropic incompressible (MEA08)~\cite{macbride_turbulent_2008}, the isotropic incompressible (PP98)~\cite{politano_von_1998} and the isotropic compressible (AS17)~\cite{andres_alternative_2017} formulations, described in detail in the Supplemental Material~\cite{supplemental_material}. The results from the three formulations are consistent within $1.3$ standard deviations [Fig.~\ref{fig:example}g]. This is a good agreement given the uncertainties; therefore, the three formulations provide a reasonable order of magnitude estimate of $|\bm{Y}|$. The energy flux $|\bm{Y}|$ shows a scaling in $l_\perp$ close to linear over a decade $\rho_i \leq l_\perp \leq l_c$, corresponding to an approximately constant energy cascade rate $\varepsilon$; this behaviour is qualitatively similar to previous \textit{in-situ} observations and numerical simulations~\cite{marino_scaling_2023}. 
This suggests that the third-order law is approximately satisfied across a limited scale range of one order of magnitude. We use the average between the three formulations $\varepsilon = (|\varepsilon_{\mathrm{PP98}}| + |\varepsilon_{\mathrm{MEA08}}| + |\varepsilon_{\mathrm{AS17}}|) / 3$ as an order of magnitude measure of the energy cascade rate. It yields, $\varepsilon\approx 8.4\pm4.2\times 10^8~\mathrm{J}~\mathrm{kg}^{-1}~\mathrm{s}^{-1}$, which is the largest energy transfer rate ever calculated from \textit{in-situ} data~\cite{osman_anisotropic_2011,hadid_compressible_2018,sorriso-valvo_turbulence-driven_2019,bandyopadhyay_observation_2020}. 

Our analysis of the statistical properties of turbulence in this example reconnection jet indicates that the energy is transferred in a spatially inhomogeneous manner across a limited inertial range. 
To our knowledge, this is the first observation of third-order laws in magnetized ($\beta_i\approx2.6$) reconnection jets. 

\begin{figure}
\includegraphics[width=\linewidth]{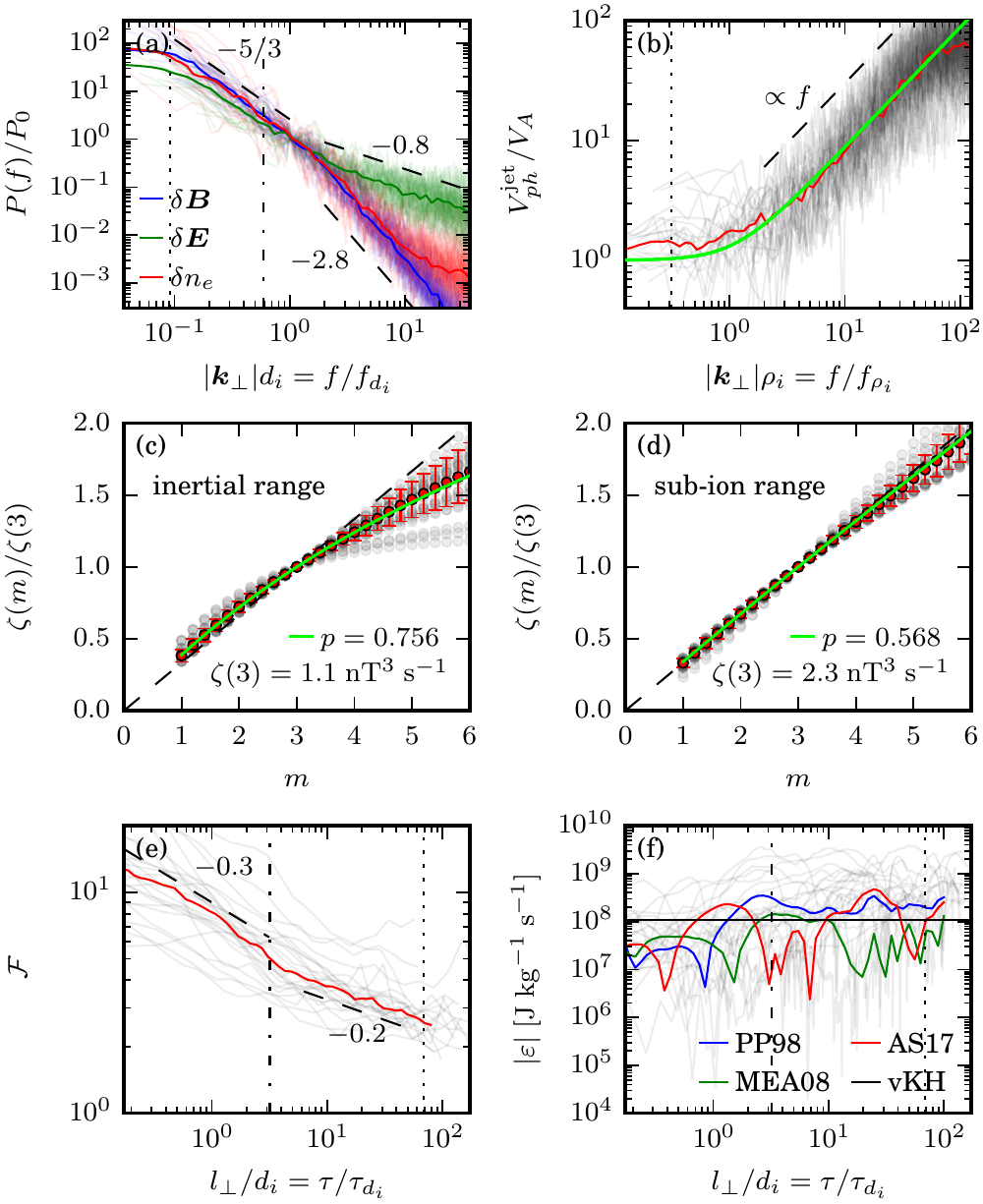}
\caption{\label{fig:sea}Superposed analysis of the 24 reconnection jets. (a) Magnetic field, electric field, and electron number density power spectra; (b) normalized phase speed in the plasma frame; (c) and (d) scaling exponents of the structure functions in the inertial and sub-ion ranges, respectively; (e) flatness; (f) energy cascade rate. The thin transparent lines show the individual rescaled cases, and the solid lines show the ensemble averages. The green lines in panels (c) and (d) are fitted $p$ model. The green line in panel (b) is the prediction for KAWs from Ref.~\cite{stasiewicz_small_2000}. The dotted lines in panels (a), (b), (e), and (f) indicate the average correlation scale $l_c$. The dashed-dotted lines in panels (a) and (e)-(f) indicate the average $|\bm{k}_\perp|\rho_i=1$ and $l_\perp=\rho_i$, respectively.}
\end{figure}

\begin{figure}
\includegraphics[width=\linewidth]{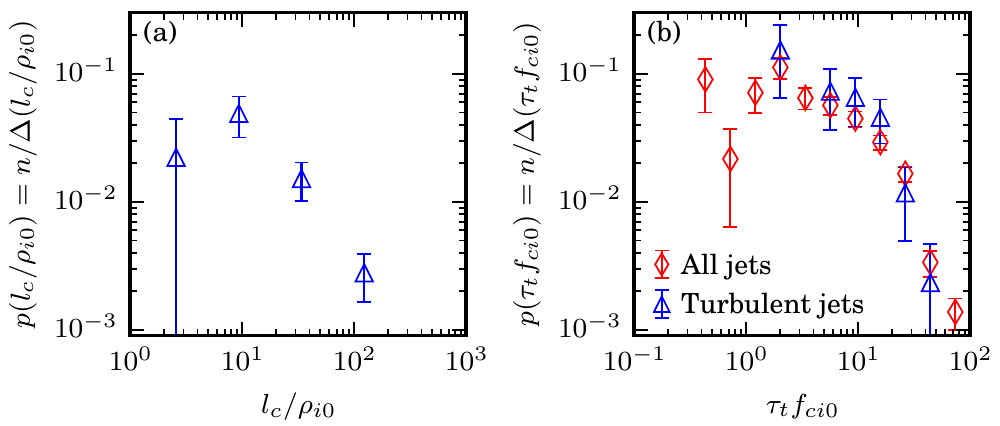}
\caption{\label{fig:histograms}Histograms of (a) the correlation scales and (b) the travel time of the jets. Blue triangles correspond to the 24 reconnection jets studied here and the red diamonds to the entire dataset.}
\end{figure}

To provide a complete systematic description of turbulence in magnetotail reconnection jets, we form an ensemble of 24 cases that show signatures of fully developed turbulence in the statistical sense introduced in the example. In the other 306 out of 330 cases, it is unclear if turbulence is developing, absent, suppressed, or statistical convergence is not achieved. 
We analyze the properties of the ensemble average of the 24 reconnection jets [Fig.~\ref{fig:sea}] after normalizing the spatial scales to the ion inertial length $d_i$ to account for the variability of the plasma conditions. This procedure results in a robust reference sample of turbulence in reconnection jets [Fig.~\ref{fig:sea}]. The ensemble-averaged magnetic field $\delta \bm{B}$, electric field $\delta \bm{E}$, and electron number density $\delta n_e$ power spectra [Fig.~\ref{fig:sea}a] show a clear power-law scaling from the injection scale $l_c$ to the ion scales, with a spectral exponent $-1.72\pm 0.03$, close to the Kolmogorov value. 
The injection scale is $l_c\sim 10\rho_{i0}$ [Fig.~\ref{fig:histograms}a] where $\rho_{i0}$ is the ion gyroradius in the background field $B_0=B_{ext}/2$~\cite{artemyev_proton_2010} with $B_{ext}=\sqrt{1 + \beta_i}|\bm{B}|$ obtained from the pressure balance assumption~\cite{asano_evolution_2003}. 
This gives $l_c\approx 3~R_E$, comparable to the typical dimension of the reconnection jet across the flow~\cite{nakamura_spatial_2004,liu_current_2013}. This suggests that turbulence in the jet is generated by its relative motion with respect to the ambient plasma.

We now examine the energy balance in the 24 intervals. Assuming that the energy injection rate in the system corresponds to the decay rate of the energy-containing eddies~\cite{matthaeus_evolution_1994}, the former can be estimated using the von K\'arm\'an-Howarth energy decay law~\cite{politano_von_1998,wan_von_2012} $\varepsilon_{\mathrm{vKH}}^\pm = -\mathrm{d} |\bm{Z}^\pm|^2/\mathrm{d}t = \alpha^\pm |\bm{Z}^\pm|^2|\bm{Z}^\mp| / l_c^\pm$, with $l_c^\pm$ the correlation length of $\bm{Z}^\pm$ and $\alpha^\pm\approx 0.03$ the von-K\'arm\'an constants~\cite{linkmann_nonuniversality_2015}. 
On the other hand, using the ensemble signed average of $\varepsilon$ from Eq.~\ref{eq:exact-law}, we estimate the energy cascade rate $\left \langle \varepsilon \right \rangle = (\left \langle \varepsilon_{\mathrm{PP98}} \right \rangle + \left \langle \varepsilon_{\mathrm{MEA08}} \right \rangle + \left \langle \varepsilon_{\mathrm{AS17}} \right \rangle) / 3 \approx 1.8_{-0.7}^{+1.1}\times 10^8~\mathrm{J}~\mathrm{kg}^{-1}~\mathrm{s}^{-1}$, positive and nearly constant at large scales, $l_c \geq l_\perp \geq \rho_i$ [Fig.~\ref{fig:sea}f]. 
The obtained value is consistent with the ensemble average total von K\'arm\'an-Howarth energy decay rate $\left \langle\varepsilon_{\mathrm{vKH}}\right \rangle=(\left \langle\varepsilon_{\mathrm{vKH}}^+\right \rangle + \left \langle\varepsilon_{\mathrm{vKH}}^-\right \rangle)/2=1.1_{-0.3}^{+1.9}\times 10^8~\mathrm{J}~\mathrm{kg}^{-1}~\mathrm{s}^{-1}$. 
This indicates that the energy injected by magnetic reconnection in the form of plasma jets is balanced by the turbulent energy transfer from the injection scale $l_c$ to the ion scales. As a result, as seen in the power spectra [Fig.~\ref{fig:sea}a], there is no energy accumulation across these scales.

To evaluate the contribution of the turbulent energy transfer to the magnetic reconnection process, we compare the energy cascade rate $\left \langle \varepsilon \right \rangle$ with the rate of decrease of magnetic energy in the reconnection inflow $\dot{\mathcal{E}}_b=(B_r^2/2\mu_0)/\Delta t$, where $B_r$ is the reconnecting magnetic field and $\Delta t$ is the duration. We note that in the approximation of Sweet-Parker reconnection, half of the energy inflow $\dot{\mathcal{E}_b}$ is available as kinetic energy in the outflow and the other half is dissipated to heating ~\cite{priest_magnetic_2000}. Assuming that $\Delta t\sim 100~\mathrm{s}$ is the typical duration of the transient reconnection~\cite{sharma_transient_2008} and $B_r=B_0\sim 10~\mathrm{nT}$, we obtain $\left \langle \varepsilon\right \rangle/\dot{\mathcal{E}}_b\sim 10\%$, suggesting that the turbulence transfers a substantial fraction of the magnetic reconnection energy input.

To understand how the energy is spatially distributed across the cascade, we analyze the high-order moments of the magnetic field fluctuations. The flatness [Fig.~\ref{fig:sea}e] monotonically increases as the scale decreases from $l_c$ to $\rho_i$, indicating a spatially inhomogeneous energy cascade at large scales. Using the structure functions $S_m(\tau)$ up to order $m=6$, we find that the scaling exponents $\zeta(m)/\zeta(3)$ [Fig.~\ref{fig:sea}c] show a non-linear monotonic increase, providing evidence of a multifractal distribution of the energy conversion sites, i.e., intermittency~\cite{frisch_turbulence_1995}. Here, we normalize the scaling exponents to $\zeta(3)=1.1 \pm 0.3$ to account for deviations from Kolmogorov's prediction $\zeta(3)=1 $~\cite{benzi_extended_1993}. To quantitatively estimate the spatial inhomogeneity of the energy cascade, we fit the observed $\zeta(m)$, with $1\leq m\leq 4$, to the multifractal $p$-model $\zeta(m)/\zeta(3) = 1 - \log_2\left [p^{m(\alpha - 1) / 2} + (1-p)^{m(\alpha - 1) / 2}\right ]$~\cite{meneveau_simple_1987}, where $p\in [0.5, 1]$ is the intermittency parameter ($p=0.5$ for mono-fractal non-intermittent fluctuations and $p=1$ for maximum intermittency) and $\alpha = \zeta(2) / \zeta(3) + 1=1.72\pm0.03$ the spectral exponent [Fig.~\ref{fig:sea}a]. We find that the normalized scaling exponents $\zeta(m)/\zeta(3)$ [Fig.~\ref{fig:sea}c] are well described by the $p$-model with $p=0.756\pm0.001$. This confirms that in the reconnection jets, the energy cascades at large scales in a multifractal, spatially inhomogeneous manner.

The statistical results described above provide evidence that an energy cascade is ongoing in this ensemble of reconnection jets at large scales. 
To quantify how fast the turbulence develops in the reconnection jets, we estimate the travel time $\tau_t$ of the jet, i.e., the time it takes for a plasma parcel to travel from the X-line to the spacecraft, which is the maximum time for turbulence to develop. Assuming an Alfv\'enic outflow, we get $\tau_t=\delta x_t / V_{A0}$, where $V_{A0}=B_0/\sqrt{\mu_0 n_i m_i}$ is the Alfv\'en speed in the background field and $\delta x_t$ is the jet travel distance between the location of the spacecraft and the statistical location of the reconnection X-line~\cite{richard_are_2022,richard_fast_2023,nagai_solar_2005}. We normalize the travel time to $f_{ci0}=eB_0/2\pi m_i$ as $\tau_t f_{ci0}=(2\pi)^{-1}\delta x_t/d_i$. 
From Fig.~\ref{fig:histograms}b, we see that the distribution of 24 reconnection jets where we find statistical signatures of fully developed turbulence mirrors that of the entire dataset of 330 cases. 
However, none of the 31 jets observed within $\tau_tf_{ci0} \leq 1$ showed signatures of fully developed turbulence. We find that the median travel time of the turbulent jets is $\tau_tf_{ci0}\approx7.2_{-3.2}^{+7.8}$, so that $\delta x_t/d_i\approx 45_{-20}^{+49}$. Development of turbulent fluctuations at similar distances ($>30 d_i$) from the reconnection X-line has been observed in simulations~\cite{higashimori_ion_2015}. Our result suggests that the turbulence can reach a well-developed state already after a few ion gyroperiods.

At scales $l_\perp\ll \rho_i$, we also observe that the scaling exponents $\zeta(m)/\zeta(3)$, with $\zeta(3)=2.3\pm 0.2$, show a weakly non-linear monotonic increase with $m$ [Fig.~\ref{fig:sea}d]. In addition, the flatness monotonically increases as the scale decreases [Fig.~\ref{fig:sea}e], indicating intermittent energy transfer at sub-ion scales. This contrasts with previous observations in other environments~\cite{wu_intermittent_2013,chen_intermittency_2014,cerri_kinetic_2019,chasapis_scaling_2020} and might be due to the growth of kinetic scale instabilities such as the ion and electron tearing modes~\cite{galeev_tearing_1976,leonardis_identification_2013} or the electron Kelvin-Helmholtz instability~\cite{che_ion_2021}. Using the multi-fractal $p$-model of energy cascade with $\alpha=2.85\pm 0.03$ yields an intermittency parameter $p=0.568\pm0.005$ close to mono-fractal ($p=0.5$). This indicates that, in contrast with the large scales, the sub-ion scale fluctuations are predominantly waves rather than turbulence.

At sub-ion scales, kinetic processes can grow into wave modes such as kinetic Alfv\'en waves (KAWs) or whistler waves. Theoretical analysis of the electron-reduced MHD~\cite{schekochihin_astrophysical_2009} and Hall MHD~\cite{galtier_wave_2006} suggested that non-linear interactions among these waves can result in an energy cascade at sub-ion scales. Kinetic-scale waves in the magnetotail plasma jets have been suggested to be KAWs~\cite{chaston_correction_2012}. To investigate the sub-ion scales energy transfer, we estimate the phase speed of the electromagnetic fluctuations in the plasma frame $V_{ph}^{\mathrm{jet}} = |\delta \bm{E}_\perp|/ |\delta \bm{B}_\perp| - V$ and compare with the prediction for KAWs~\cite{stasiewicz_small_2000,howes_astrophysical_2006,boldyrev_toward_2013} with $|\bm{k}_\perp |= |\bm{k}|$ having verified that $|\bm{k}_\perp| \gg k_\parallel$ (see Supplemental Material~\cite{supplemental_material}). The phase speed of the electromagnetic fluctuations shows a clear dispersive behaviour, $V_{ph}^{\mathrm{jet}}/V_A\propto |\bm{k}_\perp |$, in excellent agreement with the prediction for KAWs (see also Supplemental Material~\cite{supplemental_material}).

We have presented a complete systematic statistical description of turbulence in a sample of 24 reconnection jets observed by MMS. We find that the energy is injected at the characteristic scale of the jet and that turbulence transfers energy across a limited inertial range. The average zeroth-order estimate of the energy transfer rate in the MHD framework is $\left \langle \varepsilon \right \rangle = 1.8_{-0.7}^{+1.1}\times 10^8~\mathrm{J}~\mathrm{kg}^{-1}~\mathrm{s}^{-1}$, which makes reconnection jets the strongest driver of turbulence observed so far in space plasmas \cite{marino_scaling_2023}. We showed that at sub-ion scales, in contrast with the large scales, the fluctuations are weakly intermittent, indicating that they are mainly waves rather than structures. These waves are predominantly KAWs that may originate from non-linear interactions among the large-scale fluctuations or kinetic instabilities and can dissipate the energy into plasma heating through, e.g., stochastic heating and 
Landau damping~\cite{gershman_wave-particle_2017,liang_ion_2017}. 
As a result of the plasma heating, the gyroradii of the particles increase so that they can interact with the large-scale fluctuations at progressively larger scales~\cite{dalena_test-particle_2014}. Eventually, the supra-thermal particles are accelerated by the large-scale electric field of the jet~\cite{richard_proton_2022}. Thus, the jet-generated turbulence is a staircase for seed particles to climb in energy. This scenario could explain the observation of supra-thermal ion gamma-ray flares at nebula~\cite{tavani_discovery_2011} and active galactic nuclei~\cite{shukla_gamma-ray_2020}. 
Our results also provide new insights into the interplay between turbulence and magnetic reconnection. We show that reconnection outflows drive a strong turbulent cascade, which is an essential part of the fast turbulent MHD reconnection model~\cite{lazarian_turbulent_2015} and can be relevant to the generation of solar wind turbulence~\cite{zhao_turbulent_2022} by reconnection in the solar corona~\cite{drake_switchbacks_2021,raouafi_magnetic_2023}.
 
MMS data are available at the MMS Science Data Center; see Ref. \footnote{See \url{https://lasp.colorado.edu/mms/sdc/public}.}. Data analysis was performed using the \verb+pyrfu+ analysis package \footnote{See \url{https://pypi.org/project/pyrfu/}.}. 

We thank the MMS team and instrument PIs for data access and support. This work was supported by the Swedish National Space Agency (SNSA) Grants 139/18 and 145/18, and by the Swedish Research Council (VR) Research Grant 2022-03352.

\bibliographystyle{apsrev4-2}
\bibliography{arxiv}

\clearpage
\begin{widetext}
\begin{center}
\textbf{\large Supplemental Material for ``Turbulence in Magnetic Reconnection Jets\\ from Injection to Sub-Ion Scales''}
\end{center}
\end{widetext}
\setcounter{equation}{0}
\setcounter{figure}{0}
\setcounter{table}{0}
\setcounter{page}{1}
\makeatletter
\renewcommand{\theequation}{S\arabic{equation}}
\renewcommand{\thefigure}{S\arabic{figure}}

\section{1. Introduction}
We present a detailed analysis of the validity of the various assumptions used throughout the Letter. In particular, we focus on the validity of the Taylor hypothesis, the assumption of anisotropic electromagnetic fluctuations, the ergodicity theorem, and the statistical convergence. The analysis presented here for the example case in the Letter was performed on all 24 cases selected (see text in Letter). We also describe the different formulations of the third-order law employed in the Letter and estimate the contribution of the additional terms (anisotropy and compressibility) to the third-order law. Finally, we provide additional evidence that the sub-ion scale fluctuations are predominantly kinetic Alfv\'en waves (KAWs). 

\section{2. Taylor hypothesis and wavevector anisotropy}
\subsection{2.1 Taylor hypothesis}
\label{ssec:taylor}
The transformation from temporal to spatial scales using the Taylor hypothesis is crucial to compare our observations with turbulence models. 
In space plasmas, the Taylor hypothesis is satisfied if the plasma convection is much faster than the propagation of the electromagnetic field fluctuations which are effectively frozen-in-flow so that $\omega \ll |\bm{k}\cdot \bm{V_i}|$~\cite{taylor_spectrum_1938,howes_validity_2014}, i.e., $V/V_A \cos(\theta_{kV}) \gg \omega / |\bm{k}| V_A$ where $V=\left \langle |\bm{V}_i|\right \rangle$ is the flow velocity with $\bm{V}_i$ the ion bulk velocity, $V_A=|\bm{B}|/\sqrt{\mu_0 m_i n_i}$ is the Alfv\'en speed with $\bm{B}$ the magnetic field and $n_i$ the ion number density, and $\theta_{kV}=\cos^{-1}(\bm{k}\cdot\bm{V}_i/|\bm{k}||\bm{V}_i|)$ is the angle between the wavevector $\bm{k}$ and the bulk velocity. In particular, for long wavelength Alfv\'en waves with $k_\parallel\ll |\bm{k}_\perp|$ (see section 2.2), a conservative condition to satisfy the Taylor hypothesis is $V/V_A \cos(\theta_{kV}) \gg 1$. The Taylor hypothesis is known to hold in the solar wind and the Earth's magnetosheath because $V \gg V_A$. On the other hand, in the reconnection outflow, $V\leq V_A$~\cite{parker_sweets_1957,haggerty_reduction_2018}. However, observations reported that even for sub-Alfv\'enic flows, the Taylor hypothesis appears to hold~\cite{voros_bursty_2006,stawarz_observations_2016,bandyopadhyay_observation_2020}. Here, to verify the validity of the Taylor hypothesis, we apply the multi-spacecraft interferometry~\cite{graham_electrostatic_2016,graham_universality_2019} to the magnetic field ($f_{sc} < 64~\mathrm{Hz}$) with a spacecraft separation $\langle |\Delta \bm{r}|\rangle=66~\mathrm{km}=0.15d_i$. 
This technique extends the multi-spacecraft timing method~\cite{vogt_accuracy_2011} to the spectral domain using the cross-spectral density, i.e., the Fourier transform of the cross-correlation, to estimate the time delays $\Delta t_{\alpha \beta}$, where $\alpha$ and $\beta$ denote two spacecraft. 
From the obtained time delays, we can obtain the wave normal vectors, using $\Delta \bm{r}_{\alpha \beta}\cdot \hat{\bm{n}}/V_{ph}=\Delta t_{\alpha \beta}$, where $\Delta \bm{r}_{\alpha \beta}$ is the separation between the two spacecraft, $V_{ph}$ is the phase-speed, and $\hat{\bm{n}}$ is the unit wave normal vector. As a result, one can map the power spectrum $P(t, f_{sc})$ to the four-dimensional space $P(f_{sc}, k_{\perp1}, k_{\perp2}, k_{\parallel})$, where $\perp$ and $\parallel$ denote perpendicular and parallel to the background (DC) magnetic field estimated here as the mean field in the reconnection jet. 

\begin{figure}[!b]
    \centering
    \includegraphics[width=\linewidth]{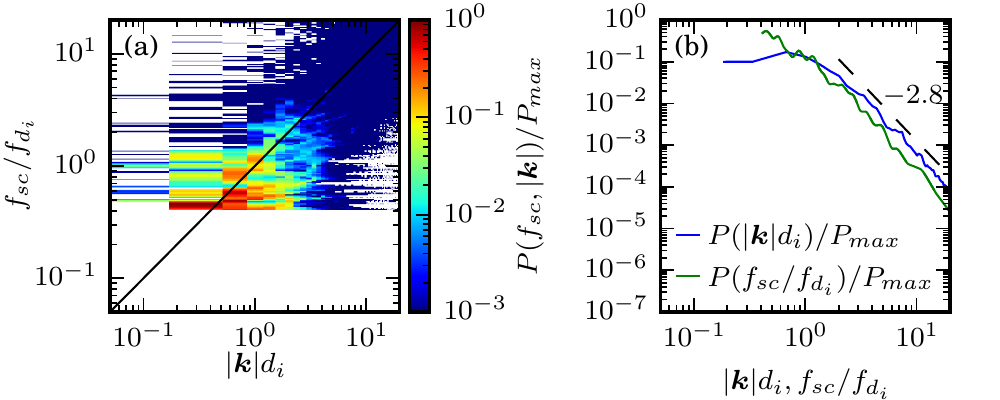}
    \caption{Joint frequency-wavenumber spectrum in the spacecraft frame. The wavenumber is normalized to the ion inertial length $d_i=\sqrt{m_i/n_i e^2 \mu_0}$ and the frequency is normalized to the Taylor transformed ion inertial length $f_{d_i}=V / 2\pi d_i$.}
    \label{fig:multispaceraft-fk}
\end{figure}

We plot, for the example shown in Fig.~1 in the Letter, the normalized magnetic field wave power as a function of the normalized spacecraft frame frequency $f_{sc}/f_{d_i}$, with $f_{d_i}=V / 2\pi d_i$, and the wavenumber $|\bm{k}|d_i$ in Figure~\ref{fig:multispaceraft-fk}a. To obtain this map, we reduced the four-dimensional space $(f_{sc}, k_{\perp1}, k_{\perp2}, k_{\parallel})$ by summing the magnetic field power for constant $|\bm{k}| = \sqrt{k_{\perp1}^2+k_{\perp2}^2+k_{\parallel}^2}$. In the limit of $\omega = 0$ and $\theta_{kV}=0$, the spacecraft frame frequency is $\omega_{sc}=|\bm{k}|V$ so that all the power should reside along the Doppler-shift line $|\bm{k}| = 2\pi f_{sc}/V$, i.e., $|\bm{k}|d_i=f_{sc}/f_{d_i}$. We observe that there is indeed a large magnetic field power near the $|\bm{k}|d_i=f_{sc}/f_{d_i}$ solid line across all scales, with some spread due to small but non-zero wave frequency $\omega\neq 0$ and non-zero angle between the wavevector and the bulk velocity $\theta_{kV}\neq0$. In addition, we show the integrated spectra $P(|\bm{k}|)=\int P(|\bm{k}|, f_{sc})\mathrm{d}f_{sc}$ and $P(f_{sc})=\int P(|\bm{k}|, f_{sc})\mathrm{d}|\bm{k}|$ in Fig.~\ref{fig:multispaceraft-fk}b. We find that the spectra computed in the wavenumber and spacecraft-frame frequency are in reasonably good agreement across the measured scales, with $(f_{sc}/f_{d_i}) / |\bm{k}|d_i\approx0.76$ minimizing the $\chi^2$ difference. We expect a better agreement at large ($|\bm{k}d_i|\leq 1$) scales, where the ions are frozen-in. We conclude that the Taylor hypothesis is sufficiently well verified from large to sub-ion scales. 

Furthermore, since we identified in the Letter that the sub-ion scale fluctuations are predominantly KAWs (see also section 6), we expect the Taylor hypothesis to be valid at these scales. Indeed, using the dispersion relation for KAWs, the condition for the validity of the Taylor hypothesis becomes $V/V_A \cos(\theta_{kV}) \gg k_\parallel d_i$~\cite{howes_validity_2014}. Since this condition is always satisfied~\cite{howes_validity_2014}, it indicates that, in our case, the Taylor hypothesis is also valid at sub-ion scales.

\subsection{2.2 Wavevector anisotropy}
\label{ssec:anisotropy}
\begin{figure}[!b]
    \centering
    \includegraphics[width=\linewidth]{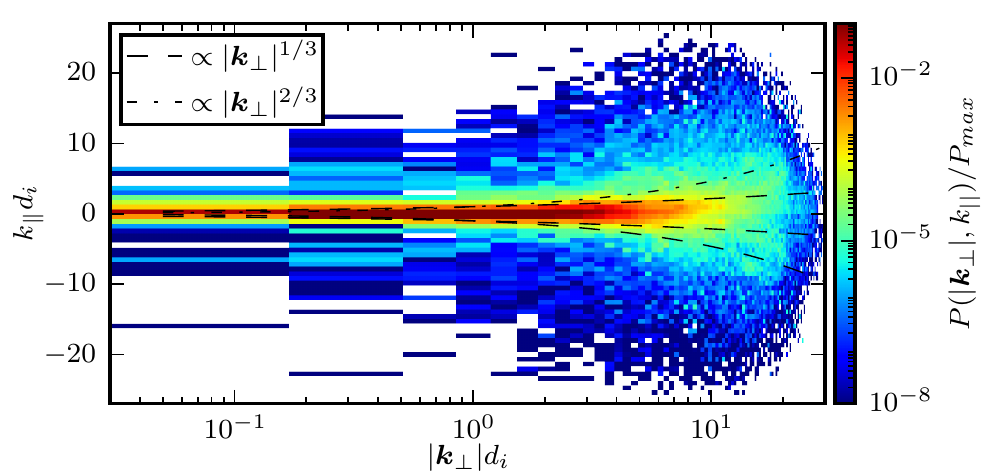}
    \caption{Magnetic field wave power in the ($|\bm{k}_\perp|d_i$, $k_\parallel d_i$) space. The dashed and dashed-dotted lines indicate $k_\parallel=|\bm{k}_\perp|^{1/3}$ and $k_\parallel=|\bm{k}_\perp|^{2/3}$, respectively.}
    \label{fig:multispaceraft-kk}
\end{figure}

To compare observations with predictions from the inertial range energy cascade ~\cite{schekochihin_astrophysical_2009} and sub-ion scale KAWs~\cite{stasiewicz_small_2000}, we used the assumption of $|\bm{k}|=|\bm{k}_\perp| \gg k_\parallel$. Turbulence is known to be strongly anisotropic in the solar wind~\cite{sahraoui_three_2010} and the Earth's magnetosheath~\cite{sahraoui_anisotropic_2006}. The $k$-filtering technique~\cite{pincon_local_1991} applied to Cluster data in a reconnection outflow showed that in the ion diffusion region $k_\parallel\gg |\bm{k}_\perp|$~\cite{eastwood_observations_2009}, while further from the reconnection region $k_\parallel\ll |\bm{k}_\perp|$~\cite{huang_observations_2012}. To verify the validity of the assumption of anisotropic turbulence, we apply the multi-spacecraft interferometry described in section 2.1.

Fig.~\ref{fig:multispaceraft-kk} presents the magnetic field wave power in the $(|\bm{k}_\perp|d_i, k_\parallel d_i)$ binned space, where $\bm{k}_\bot$ and $k_\parallel$ are the components of $\bm{k}$ perpendicular and parallel to the mean field in the reconnection jet. To obtain this map, we reduced the four-dimensional magnetic field power spectral density $P(f_{sc}, k_{\perp1}, k_{\perp2}, k_{\parallel})$ by summing the magnetic field power for constant $|\bm{k}_\bot| = \sqrt{k_{\perp1}^2+k_{\perp2}^2}$ and all frequencies $f_{sc}$. For the example presented in Fig.~1 in the Letter, the wave-power peaks for $k_\parallel\ll |\bm{k}_\perp|$, which indicates that the assumption of anisotropic electromagnetic fluctuations is valid.

\section{3. Correlation scale and ergodicity}
To estimate the correlation scale, under the Taylor frozen-in hypothesis, we use $l_c=V \tau_c$, where $l_c$ is the correlation scale, and $\tau_c=(\tau_c^+ + \tau_c^-) / 2$ the correlation time. $\tau_c^\pm$ are the $e$-folding times~\cite{smith_heating_2001} of the trace of the autocorrelation function of the Elsasser variables $R^\pm (\tau) = \left \langle \bm{Z}^\pm(t)\cdot \bm{Z}^\pm(t+\tau)\right \rangle_T$, with $\left \langle \cdot\right \rangle_T$ denoting the ensemble time average and $\bm{Z}^\pm = \bm{V_i}\pm \bm{B}/\sqrt{\mu_0 m_i n_i}$ are the Elsasser fields. 

\begin{figure}[!b]
    \centering
    \includegraphics[width=\linewidth]{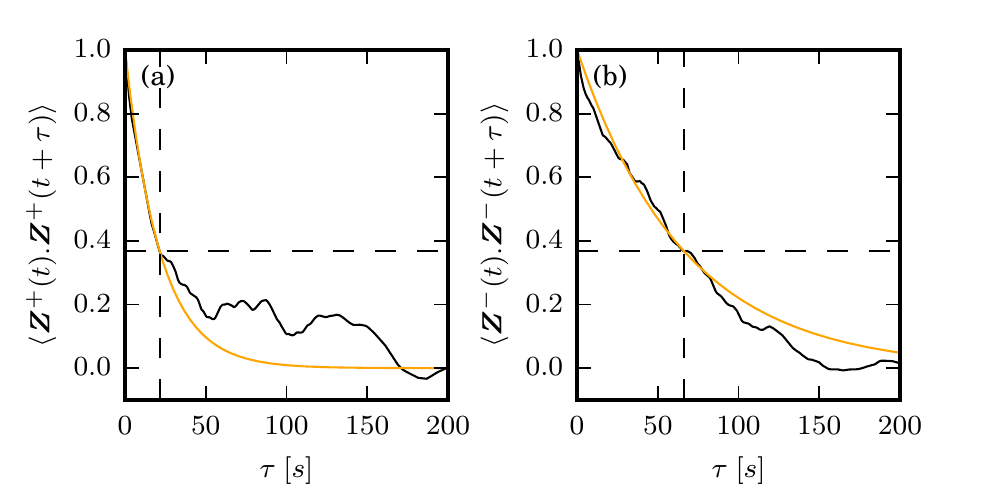}
    \caption{Autocorrelation function of (a) $\bm{Z}^+$ and (b) $\bm{Z}^-$. The dashed lines indicate the $e$-folding time, and the orange line the corresponding decaying exponential.}
    \label{fig:autocorr}
\end{figure}

Fig.~\ref{fig:autocorr} presents the autocorrelation function of the Elsasser variables $\bm{Z}^\pm$ for the example [Fig.~1 in the Letter]. We observe that the autocorrelation function is well-fitted by a decaying exponential. In particular, $\tau_c=44~\mathrm{s}$, with $\tau_c^+=22~\mathrm{s}$ and $\tau_c^-=66~\mathrm{s}$. 
The reconnection jet interval 2017-05-28T00:35:26.553 - 2017-05-28T00:38:58.054 UT contains $4.8~\tau_c$. Hence, the ergodicity theorem is satisfied.

\section{4. Statistical convergence}
To provide a reliable statistical description of the turbulence, we must first ensure convergence of the moments of the probability distribution function (PDF) of the magnetic field and velocity increments $\Delta \bm{Z}^\pm (\tau) = \bm{Z}^\pm (t+\tau) - \bm{Z}^\pm (t)$. We verified the convergence of the PDF moments against several tests~\cite{dudok_de_wit_can_2004,kiyani_extracting_2006}. Here, we present the more restrictive test we used. 

 One can show that for a finite sample size, the $m$th order moment of the increments diverges if $m\gamma>1$ with $\gamma$ the scaling index of the ranked distribution of $\Delta \bm{Z}^\pm$~\cite{dudok_de_wit_can_2004}. Hence, moments of the PDF of the increments of the Elsasser variables are only meaningful up to the order~\cite{dudok_de_wit_can_2004}
 
\begin{equation}
    m_{max}=\left \lfloor\frac{1}{\gamma}\right \rfloor - 1 \, .
\end{equation}

\begin{figure}[!t]
    \centering
    \includegraphics[width=\linewidth]{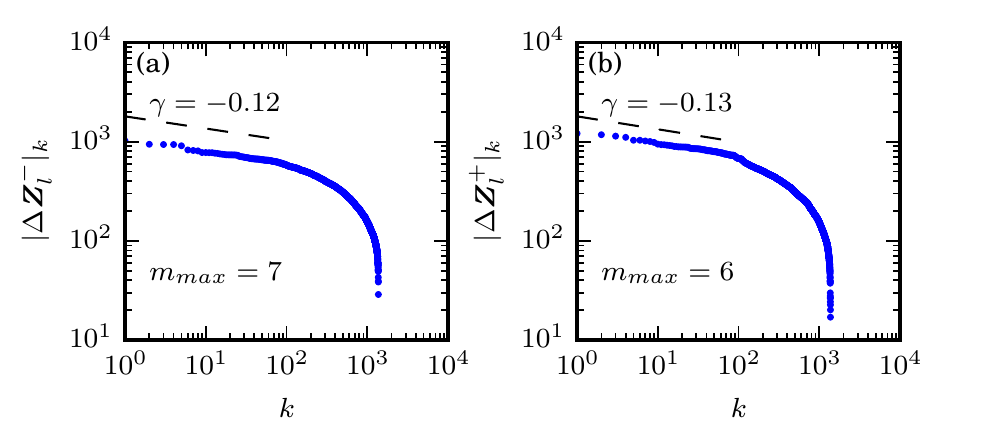}
    \caption{Statistical convergence test following~\cite{dudok_de_wit_can_2004}. Ranked increments (a) $\Delta \bm{Z}^-_l$ and (b) $\Delta \bm{Z}^+_l$.}
    \label{fig:convergence}
\end{figure}

We compute $\Delta \bm{Z}^\pm (\tau)$ for $\tau=V /d_i$. We see that the ranked distribution of  $\Delta \bm{Z}^\pm (\tau)$ behaves as a power law up to $10^2$ [Fig.~\ref{fig:convergence}]. In particular, a fit with the Levenberg-Marquart least square fitting method yields $\gamma=0.125\pm 0.002$ for  $\Delta \bm{Z}^-$ and $\gamma=0.132\pm 0.002$ for  $\Delta \bm{Z}^+$. Hence, $m_{max} = 6$, which is the maximum order of moments that can be meaningfully assessed (e.g., to estimate $p$ from the multi-fractal $p$-model~\cite{meneveau_simple_1987}). 

\section{5. Formulations of the third-order law}
As described in the Letter, the third-order law for three-dimensional single-fluid magnetohydrodynamic (MHD) can be simplified to formulations which can be applied to spacecraft measurements. Here, we provide a detailed description of the three formulations employed in the Letter.

\subsection{5.1 Isotropic incompressible}
Under the assumption of isotropy and incompressibility of the turbulence, Ref.~\cite{politano_von_1998} showed that the third-order law [Eq. 1 in the Letter] can be re-written as

\begin{equation}
    \label{eq:pp98}
    Y^{\pm} = \left \langle |\Delta \bm{Z}^\pm (t, \tau)|^2 \Delta Z_l^\mp(t, \tau)\right \rangle = -\frac{4}{3} \varepsilon^\pm V \tau \, ,
\end{equation}
\noindent
where $\Delta Z_l^\mp(t, \tau) = \Delta \bm{Z}^\mp (t, \tau)\cdot\bm{V}_i$, and $\varepsilon^\pm$ are the associated energy cascade rates. 

\begin{figure}[!b]
    \centering
    \includegraphics[width=\linewidth]{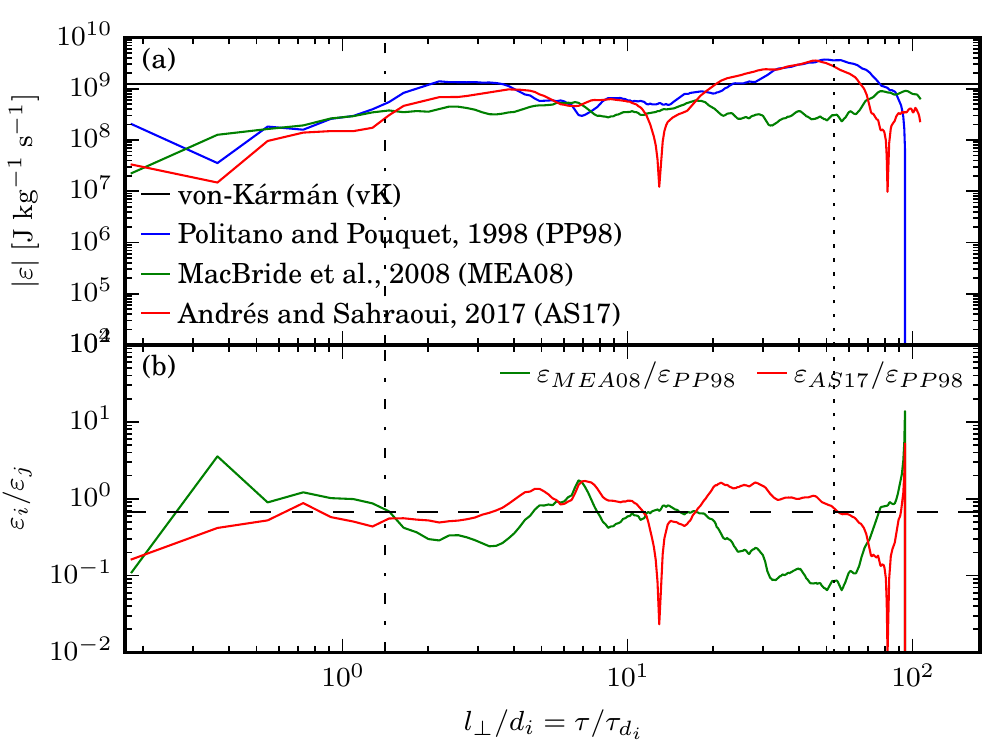}
    \caption{\label{fig:examplepp98mea08as17}Comparison between the different formulations of the third-order law for the example reconnection jet. (a) Energy cascade rate estimated using: von K\'arm\'an-Howarth energy decay rate (black), isotropic incompressible MHD~\cite{politano_von_1998} (blue), anisotropic incompressible MHD~\cite{macbride_turbulent_2008} (green), and isotropic compressible MHD~\cite{andres_alternative_2017} (red), (b) Ratio of the energy cascade rate from the different formulations. The dotted and dashed-dotted lines indicate the correlation scale and the ion gyroradius, respectively.}
\end{figure}

We plot the incompressible isotropic energy transfer rate for the example event presented in Fig. 1 in the Letter and the 24 reconnection jets in Figures~\ref{fig:examplepp98mea08as17} and~\ref{fig:statspp98mea08as17}, respectively. For the example and the 24 cases, the energy cascade rate is nearly constant across one decade $(\rho_i\leq l_\perp \leq l_c)$. 

\subsection{5.2 Anisotropic incompressible}
\begin{figure}[!b]
    \centering
    \includegraphics[width=\linewidth]{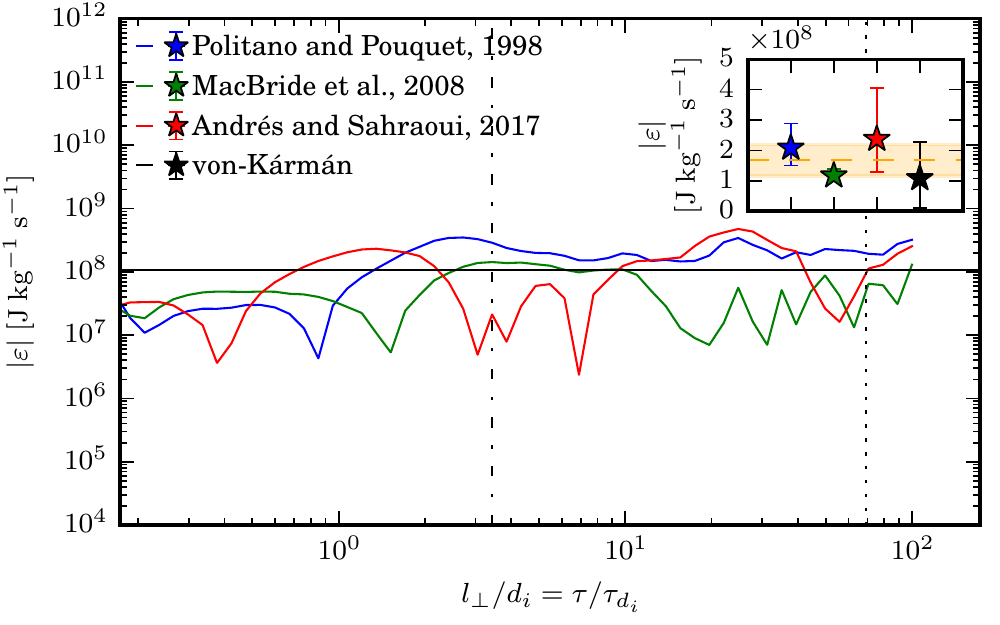}
    \caption{\label{fig:statspp98mea08as17}Comparison between the different formulations of the third-order law: von K\'arm\'an-Howarth energy decay rate (black), isotropic incompressible MHD~\cite{politano_von_1998} (blue), anisotropic incompressible MHD~\cite{macbride_turbulent_2008} (green), and isotropic compressible MHD~\cite{andres_alternative_2017} (red) for the ensemble of 24 reconnection jets. The solid lines correspond to $\varepsilon>0$ and the dashed lines to $\varepsilon<0$. The dotted and dashed-dotted lines indicate the correlation scale and the ion gyroradius, respectively.}
\end{figure}

As we discussed above (see section 2.2), the turbulence in the reconnection jets is strongly anisotropic, which suggests that an additional contribution ignored in the isotropic formulation of the third-order law [Eq.~\ref{eq:pp98}] can affect the turbulent energy transfer~\cite{macbride_turbulent_2008,stawarz_turbulent_2009}. We employ the hybrid approach formulated in Ref.~\cite{macbride_turbulent_2008}. This approach consists of projecting the third-order law onto a 2D slice perpendicular to the mean magnetic field and a 1D component parallel to the mean magnetic field. To do so we use the mean-field coordinates defined as $\bm{\hat{e}}_{\perp 1}=(\bm{\hat{e}}_V\times\bm{\hat{e}}_B)/|\bm{\hat{e}}_V\times\bm{\hat{e}}_B|$, $\bm{\hat{e}}_{\perp 2}=\bm{\hat{e}}_{\parallel}\times\bm{\hat{e}}_{\perp 1}$, and $\bm{\hat{e}}_{\parallel}=\bm{\hat{e}}_B$, where $\bm{\hat{e}}_V=\bm{V}_i/|\bm{V}_i|$ and $\bm{\hat{e}}_B=\bm{B}/|\bm{B}|$~\cite{bieber_dominant_1996}. In this formulation, the total energy transfer rate reads $\varepsilon^\pm=\varepsilon_{\perp}^\pm/2 + \varepsilon_{\parallel}^\pm/4$, where $\varepsilon_{\perp}$ and $\varepsilon_{\parallel}$ are defined as 

\begin{eqnarray}
    Y_{\perp}^\pm(t,~\tau) & = & \left \langle \left |\Delta \bm{Z}^\pm(t,~\tau) \right |^2 \Delta Z_{\perp 2}^\mp(t,~\tau) \right \rangle  \nonumber \\
     & = & 2\varepsilon_{\perp}^\pm V \tau \sin(\theta_{BV})\, ,
\end{eqnarray}
and 
\begin{eqnarray}
    Y_{\parallel}^\pm(t,~\tau) & = & \left \langle \left |\Delta \bm{Z}^\pm(t,~\tau) \right |^2 \Delta Z_{\parallel}^\mp(t,~\tau) \right \rangle \nonumber \\
     & = & 4\varepsilon_{\parallel}^\pm V \tau \cos(\theta_{BV})\, ,
\end{eqnarray}

\noindent
with $\Delta Z_{\perp 2} = \Delta \bm{Z}\cdot\bm{\hat{e}}_{\perp 2}$ and $\Delta Z_{\parallel} = \Delta \bm{Z}\cdot\bm{\hat{e}}_{\parallel}$ and $\theta_{BV}=\cos^{-1}(\bm{\hat{e}}_B\cdot\bm{\hat{e}}_V)$ is the angle between the velocity and the mean magnetic field. We plot $\varepsilon$ estimated using the isotropic and the anisotropic formulations for the example event and for the 24 reconnection jets in Figs.~\ref{fig:examplepp98mea08as17} and~\ref{fig:statspp98mea08as17}, respectively.

For the example, we observe that similar to the isotropic formulation, the scaling is indeed very good, in the sense that $\varepsilon$ is nearly constant across the large scales ($l_c\geq l_\perp \geq \rho_i$). The estimate of the energy transfer rate using the anisotropic formulation is in reasonable agreement with that obtained using the isotropic formulation ($\varepsilon_{MEA08}/\varepsilon_{PP98} = 0.52$) [Fig.~\ref{fig:examplepp98mea08as17}]. For the ensemble of 24 events, the scaling is also very good over one order of magnitude, and, similar to the example, the estimated energy transfer rate is in reasonable agreement (within $1.3$ standard deviations) [Fig.~\ref{fig:statspp98mea08as17} inset] with that obtained using the isotropic formulation of the third-order law. Given the numerous potential sources of error that can affect the estimate of the energy transfer rate (e.g., uncertainty in plasma measurements, the method itself~\cite{wang_strategies_2022}, the incompleteness of the formulation, the statistical variability of the conditions, etc.), this indicates that the estimate of the energy transfer rate provided in the Letter is reliable to the zeroth order.

\subsection{5.3 Isotropic compressible}
In the above-described formulations, we assume that the turbulence is incompressible to compute the energy cascade rate $\varepsilon$. However, in the Letter, we demonstrate that the fluctuations are predominantly KAWs at sub-ion scales, which are compressible~\cite{schekochihin_astrophysical_2009,boldyrev_toward_2013}. Since it was shown that the contribution of the compressible effects to the energy cascade rate could be comparable to the incompressible part~\cite{andres_energy_2019}, we estimate the contribution of the compressible effects to the energy cascade rate in our sample of reconnection jets using the isotropic compressible MHD theory~\cite{andres_alternative_2017}. Neglecting the contribution of the energy source terms with respect to the flux terms~\cite{andres_energy_2019}, the isotropic compressible third-order law is given by

\begin{eqnarray}
-\frac{4}{3}\varepsilon^\pm V \tau \rho_0 & = & \left \langle \left [\Delta (\rho \bm{Z}^\pm) \cdot \Delta \bm{Z}^\pm \right ] \Delta Z_l^\mp\right \rangle \nonumber \\
 & & + \left \langle \Delta \rho \Delta u \Delta V_l\right \rangle \, ,
\end{eqnarray}

\noindent
where $\rho = \rho_0 + \delta \rho$ is the local plasma density with $\rho_0=\left \langle \rho \right \rangle$ and $u=C_s^2\log(\rho/\rho_0)$ is the internal energy with $C_s^2=\beta V_A^2/2$ the isothermal ion sound speed. We plot $\varepsilon$ estimated using incompressible and compressible formulations for the example event presented in Fig. 1 in the Letter and the 24 reconnection jets in Figs.~\ref{fig:examplepp98mea08as17} and~\ref{fig:statspp98mea08as17}, respectively.

For the example, we find that the energy transfer rate obtained considering the compressibility effects is almost equal to that obtained in the incompressible MHD framework. The energy cascade rate estimated using the two formulations is similar, with $\varepsilon_{AS17}/\varepsilon_{PP98} = 0.82$. On the other hand, for the ensemble of 24 reconnection jets, we observe that the scaling in the compressible MHD framework is not as satisfactory as the incompressible part. This can be attributed to either the absence of statistical convergence when including the density or missing additional compressible terms not included in this formulation~\cite{simon_exact_2022}. Nevertheless, we find that the estimated compressible energy cascade rate is in very good agreement with the incompressible energy cascade rate [Fig.~\ref{fig:statspp98mea08as17} inset], which indicates that the estimate of the energy transfer rate provided in the Letter is reliable to the zeroth order.

\section{6. Identification of the nature of sub-ion scale fluctuations}
In the Letter, we show that the plasma frame phase speed of the sub-ion scale fluctuations is consistent with the prediction for KAWs. To provide additional evidence on the nature of the sub-ion scale fluctuations, we investigate the magnetic and plasma compressibility~\cite{gary_short-wavelength_2009,chen_nature_2013,groselj_fully_2018}. At sub-ion scales, the kinetic Alfv\'en and the whistler modes are the two relevant electromagnetic modes~\cite{schekochihin_astrophysical_2009,boldyrev_toward_2013}. The major difference between these two modes is that, for the observed plasma conditions, the whistler mode is nearly incompressible while the kinetic Alfv\'en mode is compressible~\cite{schekochihin_astrophysical_2009,gary_short-wavelength_2009,boldyrev_toward_2013}.

For the whistler mode to not be significantly damped compared with the kinetic Alfv\'en mode, the ion contribution (dissipation) $\gamma_i=-2\sqrt{\pi}\beta_i^{-3/2} k_\parallel d_i \exp \left (-k_\parallel^2d_i^2/\beta_i\right )$ to the growth rate must remain small~\cite{boldyrev_toward_2013}. Using an $e$-folding threshold for the observed $\beta_i\approx 2.6$, $k_\parallel d_i$ must satisfy $k_\parallel d_i \gtrsim 1.987$. As a result, the electron compressibility is such that $C_e=\left (\delta n_e/n_{e0}\right )^2/\left (|\delta \bm{B}|/|\bm{B}_0|^2\right ) \lesssim 0.03$. On the other hand, for KAWs $C_e= 2/\beta \left (\beta + 1\right )$ and $C_\parallel=\delta B_\parallel^2/|\delta \bm{B}|^2 = \beta/2\left (1 + \beta\right )$~\cite{boldyrev_toward_2013}, with $\beta=\beta_i+\beta_e$. Here, we obtain $C_e=0.36_{-0.13}^{+0.12}$, an order of magnitude larger than for whistler waves, providing a clear distinction between the two modes.

\begin{figure}[!b]
    \centering
    \includegraphics[width=\linewidth]{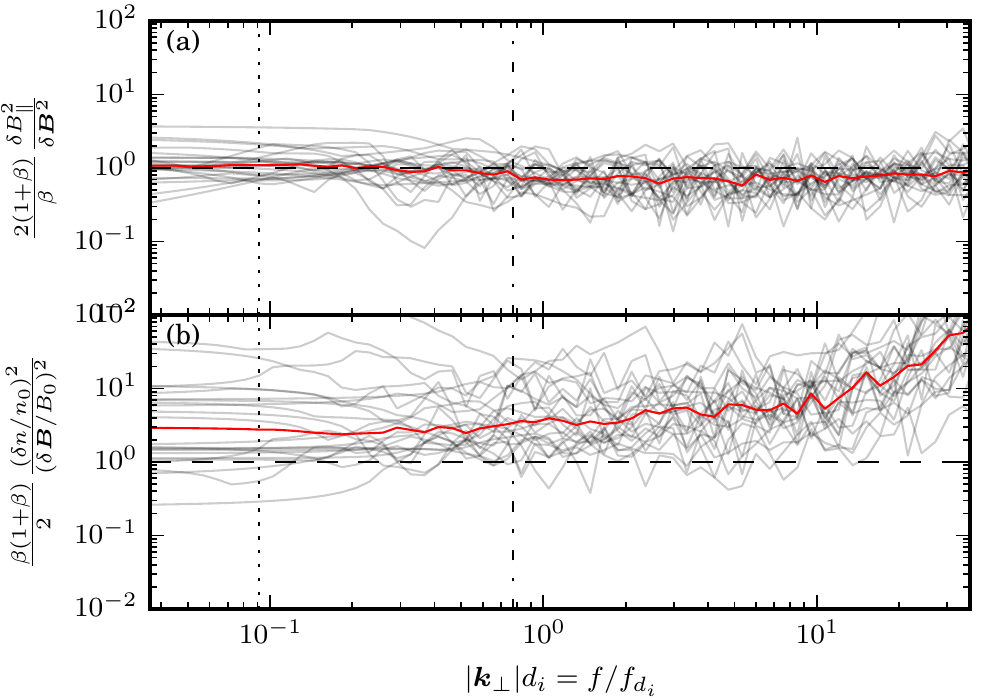}
    \caption{\label{fig:kaw}Normalized magnetic (a) and electron (b) compressibility for the 24 reconnection jets. The red line indicated the median, and the black dashed the theoretical prediction for KAWs~\cite{boldyrev_toward_2013}. The dotted and dashed-dotted lines indicate the correlation scale and the ion gyroradius, respectively.}
\end{figure}

We plot the normalized magnetic compressibility $\tilde{C}_\parallel=\left [2(1+\beta)/\beta\right ] C_\parallel$ [Fig.~\ref{fig:kaw}a] and the normalized electron compressibility $\tilde{C}_e= \left [\beta(1+\beta)/2\right ]C_e$ [Fig.~\ref{fig:kaw}b]. We find that $\tilde{C}_\parallel\sim 1$ and $\tilde{C}_e\sim 1$ agree with theoretical predictions for KAWs. This provides additional evidence that the fluctuations are predominantly KAWs at sub-ion scales.

\end{document}